\documentclass[12pt]{article}

\usepackage{graphicx}

\begin{document}

\begin{center}
{\bf Universe acceleration and nonlinear electrodynamics } \\
\vspace{5mm} S. I. Kruglov
\footnote{serguei.krouglov@utoronto.ca}

\vspace{3mm}
\textit{Department of Chemical and Physical Sciences, University of Toronto,\\
3359 Mississauga Road North, Mississauga, Ontario L5L 1C6, Canada} \\
\vspace{5mm}
\end{center}

\begin{abstract}
A new model of nonlinear electrodynamics with a dimensional parameter $\beta$ coupled to gravity is considered.
We show that an accelerated expansion of the universe takes place if the nonlinear electromagnetic field is the source of the gravitational field. A pure magnetic universe is investigated and the magnetic field drives the universe to accelerate. In this model, after the big bang, the universe undergoes inflation, and the accelerated expansion and then decelerates approaching Minkowski spacetime asymptotically. We demonstrate the causality of the model and a classical stability at the deceleration phase.
\end{abstract}

\section{Introduction}

An accelerated expansion of the universe was confirmed by observations of the redshift of type Ia supernovae
and the cosmic microwave background (CMB). To explain the acceleration of the universe one can introduce the cosmological constant, $\Lambda$, in the Einstein equation with the unusual equation of state $p=-\rho$ ($p$, $\rho$ are the pressure and the energy density, respectively). The fluid with such a property is called dark energy (DE).
From the theoretical point of view, it is difficult to make such an introduction of $\Lambda$ clear.
Another way to describe DE is to consider a scalar field (quintessence) with the proper potential which drives the
universe to accelerate. The nature of the scalar field and its potential are also not easily understood.
There are attempts to modify gravity by the replacement of the Ricci scalar $R$ in the Einstein-Hilbert (EH) action by some function $F(R)$ [$F(R)$ gravity models] \cite{Capozziello} to explain the universe acceleration. There are many $F(R)$ gravity models and there are not derivations of definite functions $F(R)$ from first principles.
At the same time cosmological models exploring nonlinear electrodynamics (NLE) may solve problems of early time inflation and singularities without the modification of General Relativity (GR). Some NLE models,
which are effective models that take into account quantum corrections,
do not have divergences and possess finite self-energy of charge particles. Some examples are the Maxwell Lagrangian plus
the Heisenberg-Euler Lagrangian \cite{Heisenberg} taking into account one-loop QED quantum corrections and the Born-Infeld theory \cite{Born}, that can be used for strong electromagnetic fields, smoothing divergences. The electromagnetic and gravitational fields in the early epoch are very strong, and therefore, the nonlinear electromagnetic effects should be taken into consideration. NLE models can mimic DE near the Planck era and give inflation of the early universe. The magnetic fields in the early universe epoch may be in the order of $10^{15}$ G and classical electrodynamics has to be modified \cite{Jackson}. The Maxwell theory can be considered as an approximation of NLE theory for weak fields and for strong fields we should use spacetimes with NLE fields. Thus, NLE has been used to produce inflation in the early universe \cite{Garcia}, \cite{Camara}, and it is a source of the universe acceleration \cite{Elizalde} - \cite{Novello1}. The coupling NLE with gravity may give negative pressure that drives the universe to accelerate \cite{Novello} - \cite{Vollick}.
In this paper, we explore a new NLE model \cite{Kruglov} so that the electromagnetic field is coupled to the gravitational field. We investigate the evolution of the universe in the Einstein-NLE model which prevents the initial big bang singularity. GR can be treated as an effective field theory at low energy of quantum gravity theory, and the EH classical action should have the trace anomaly of the energy-momentum tensor \cite{Mottola}. The result of breaking the scale invariance in NLE gives the negative pressure.

We use units with $c=\hbar=\varepsilon_0=\mu_0=1$ and the metric $\eta=\mbox{diag}(-,+,+,+)$. Greek letters run $0, 1, 2, 3$ and Latin letters run $1, 2, 3$.

\section{Nonlinear electrodynamics and cosmology}

Electromagnetic fields in cosmology are of great interest because of the CMB observed.
Here we consider the effective theory of nonlinear electromagnetic fields with the dimensional parameter $\beta$ suggested in Ref. \cite{Kruglov} and its influence on cosmology. The parameter $\beta^{1/4}$ possesses the dimension of the length,  and there is the maximum of the possible electric field strength $E_{max}=1/\sqrt{\beta}$ in the model under consideration. The electric field of a pointlike charged particle does not possess singularity at the origin, and
the electromagnetic energy of a pointlike particle is finite. In this model the mass of the charged particle can be treated as pure electromagnetic energy.
One can speculate that the parameter $\beta^{1/4}$ is connected with the fundamental length going from quantum gravity. The Lagrangian density of NLE introduced in Ref. \cite{Kruglov} is given by
\begin{equation}
{\cal L}_{em} = -\frac{{\cal F}}{2\beta{\cal F}+1},
\label{1}
\end{equation}
where $\beta{\cal F}$ is dimensionless, ${\cal F}=(1/4)F_{\mu\nu}F^{\mu\nu}=(\textbf{B}^2-\textbf{E}^2)/2$, and $F_{\mu\nu}$ is the field strength tensor. The Lagrangian (1) would not have a denominator vanishing at some point because the electric field strength cannot reach the value $E_{max}=1/\sqrt{\beta}$.
The energy-momentum tensor was obtained in Ref. \cite{Kruglov} and is
\begin{equation}
T_{\mu\nu}=-\frac{1}{(2\beta{\cal F}+1)^2}\left[F_{\mu}^{~\alpha}F_{\nu\alpha}-g_{\mu\nu}{\cal F}(2\beta{\cal F}+1)\right],
 \label{2}
\end{equation}
and it has a nonvanishing trace. The scale invariance in the NLE model based on Eq. (1) is broken, and that allows for the negative pressure.
We should make the average of the electromagnetic fields which are sources in GR \cite{Tolman} to have the isotropy of the Friedman-Robertson-Walker (FRW) spacetime. Thus, we use the average values of the electromagnetic fields as follows:
\[
<\textbf{E}>=<\textbf{B}>=0,~~~~<E_iB_j>=0,
\]
\begin{equation}
<E_iE_j>=\frac{1}{3}E^2g_{ij},~~~~<B_iB_j>=\frac{1}{3}B^2g_{ij}.
\label{3}
\end{equation}
We consider an average over a volume which is large as compared to the radiation
wavelength, and it is small compared to the spacetime curvature. In the following we will omit brackets for simplicity.
From Eq. (2), we obtain the density energy $\rho$ and the pressure $p$,
\begin{equation}
\rho=\frac{E^2}{\left(2\beta{\cal F}+1\right)^2}+\frac{{\cal F}}{2\beta{\cal F}+1},
\label{4}
\end{equation}
\begin{equation}
p=\frac{2B^2-E^2}{3\left(2\beta{\cal F}+1\right)^2}-\frac{{\cal F}}{2\beta{\cal F}+1}.
\label{5}
\end{equation}
The action of GR coupled with the nonlinear electromagnetic field described by the Lagrangian density (1) is
\begin{equation}
S=\int d^4x\sqrt{-g}\left[\frac{1}{2\kappa^2}R+ {\cal L}_{em}\right],
\label{6}
\end{equation}
where $\kappa^{-1}=M_{Pl}$, $M_{Pl}$ is the reduced Planck mass, and $R$ is the Ricci scalar.
The Einstein and electromagnetic field equations follow from Eq. (6),
\begin{equation}
R_{\mu\nu}-\frac{1}{2}g_{\mu\nu}R=-\kappa^2T_{\mu\nu},
\label{7}
\end{equation}
\begin{equation}
\partial_\mu\left(\frac{\sqrt{-g}F^{\mu\nu}}{(2\beta{\cal F}+1)^2}\right)=0.
\label{8}
\end{equation}
With the help of the FRW metric and the Einstein equation (7), one obtains Friedmann's equation as follows,
\begin{equation}
3\frac{\ddot{a}}{a}=-\frac{\kappa^2}{2}\left(\rho+3p\right),
\label{9}
\end{equation}
where $a$ is a scale factor and dots over the $a$ mean the derivatives with respect to the cosmic time.
The inequality $\rho + 3p < 0$ is needed in order to have an accelerated universe. Here we suppose that the field of NLE is the main source of gravity. We consider the case when $\textbf{E} = 0$ because only the magnetic field is important in cosmology.
Thus, the electric field is screened because of the charged primordial plasma, but the magnetic field is not screened \cite{Lemoine}. In accordance with the standard cosmological model, there is no asymmetry in the direction, and this is satisfied by the requirement $<B_i> = 0$. As a result, the magnetic field does not induce the directional effects.
From Eqs. (4) and (5), we find
\begin{equation}
\rho+3p=\frac{B^2\left(1-\beta B^2\right)}{\left(\beta B^2+1\right)^2}.
\label{10}
\end{equation}
So, the requirement $\rho + 3p < 0$ for the accelerating universe is satisfied if $\beta B^2>1$. At the early epoch with strong magnetic fields, this requirement can occur. Therefore, in NLE under consideration,
the inequality $\rho + 3p < 0$ can be satisfied, and the magnetic field under the conditions (3) drives the accelerated expansion of the universe.
From the conservation of the energy-momentum tensor, $\nabla^\mu T_{\mu\nu}=0$, for the FRW metric, one obtains
\begin{equation}
\dot{\rho}+3\frac{\dot{a}}{a}\left(\rho+p\right)=0.
\label{11}
\end{equation}
From Eqs. (4) and (5), for the case $E^2 = 0$, we find
\begin{equation}
\rho=\frac{B^2}{2\left(\beta B^2+1\right)},~~~~\rho+p=\frac{2B^2}{3\left(\beta B^2+1\right)^2}.
\label{12}
\end{equation}
Replacing $\rho$ and $p$ from Eqs. (12) into Eq. (11) and integrating, one obtains
\begin{equation}
B(t)=\frac{B_0}{a(t)^2}.
\label{13}
\end{equation}
Equation (13) shows the evolution of the magnetic field with the scale factor. From Eq. (12), we find the dependence of the energy density and the pressure on the scale factor
\begin{equation}
\rho(t)=\frac{B_0^2}{2\left[\beta B_0^2+a(t)^4\right]},~~~~p(t)=\frac{B_0^2a(t)^4-3\beta B_0^4}{6\left[\beta B_0^2+a(t)^4\right]^2}.
\label{14}
\end{equation}
It follows from Eqs. (14) that there is not singularity of the energy density nor the pressure at $a(t)\rightarrow 0$ and $a(t)\rightarrow \infty$. Thus,
\begin{equation}
\lim_{a(t)\rightarrow 0}\rho(t)=\frac{1}{2\beta},~~\lim_{a(t)\rightarrow 0}p(t)=-\frac{1}{2\beta},~~\lim_{a(t)\rightarrow \infty}\rho(t)=\lim_{a(t)\rightarrow \infty}p(t)=0.
\label{15}
\end{equation}
Equations (15) show that at the beginning of the universe evolution ($a=0$) the model gives $\rho=-p$, i.e. the same property as in the $\Lambda$CDM model.
The absence of singularities is an attractive feature which is connected with the particular NLE \cite{Kruglov}.
From Einstein's equation (7) and the energy-momentum tensor (2) we obtain the Ricci scalar
\begin{equation}
R=\kappa^2T_{\mu}^{~\mu}=\frac{8\beta \kappa^2{\cal F}^2}{(1+2\beta {\cal F})^2}.
\label{16}
\end{equation}
Taking into account Eq. (13) one finds the evolution of the curvature
\begin{equation}
R(t)=\frac{2\beta \kappa^2B_0^4}{\left[a(t)^4+\beta B_0^2\right]^2}=\kappa^2(\rho-3p).
\label{17}
\end{equation}
One can verify that expression (17) coincides with the Ricci scalar given in Eq. (29) (in the Appendix).
We obtain from Eqs. (15),(17),(34), and (35) (see the Appendix)
\[
\lim_{a(t)\rightarrow 0}R(t)=\frac{2\kappa^2}{\beta},~~~~\lim_{a(t)\rightarrow \infty}R(t)=0.
\]
\[
\lim_{a(t)\rightarrow 0}R_{\mu\nu}R^{\mu\nu}=\frac{\kappa^4}{\beta^2},
~~~~\lim_{a(t)\rightarrow \infty}R_{\mu\nu}R^{\mu\nu}=0.
\]
\begin{equation}
\lim_{a(t)\rightarrow 0}R_{\mu\nu\alpha\beta}R^{\mu\nu\alpha\beta}=\frac{2\kappa^4}{3\beta^2},~~~~
\lim_{a(t)\rightarrow \infty} R_{\mu\nu\alpha\beta}R^{\mu\nu\alpha\beta}=0.
\label{18}
\end{equation}
As a result, there are not singularities of the Ricci scalar, the Ricci tensor squared $R_{\mu\nu}R^{\mu\nu}$, nor the Kretschmann scalar $R_{\mu\nu\alpha\beta}R^{\mu\nu\alpha\beta}$ at $a(t)\rightarrow 0$ and $a(t)\rightarrow \infty$. The scale factor increases in the time and at the infinite time spacetime becomes flat (Minkowski spacetime), and there is not singularity in the future. It follows from Eqs. (10) and(13) that the universe accelerates at $a(t)<\beta^{1/4}\sqrt{B_0}$. So, the universe acceleration at the early time can be explained in the model under consideration.

\section{Cosmic evolution}

We will study the dynamics of the universe with the energy density (14) and neglect the dustlike matter using  Einstein's equations. We make a conclusion from Eq. (9) that at the critical value ($a_c=\beta^{1/4}\sqrt{B_0}$) of the scale factor, $a=a_c$, the universe acceleration is zero. The evolution of the scale factor can be studied by analyzing the second Friedmann equation without the cosmological constant
\begin{equation}
\left(\frac{\dot{a}}{a}\right)^2=\frac{\kappa^2\rho}{3}-\frac{\epsilon}{a^2},
\label{19}
\end{equation}
where $\epsilon=0$ corresponds to three-dimensional flat universe and $\epsilon=1$ - for closed and $\epsilon=-1$ - for open universes, respectively. Equation (19) may be represented with the help of Eq. (14) as
\begin{equation}
\dot{a}^2 + V(a)=-\epsilon,
\label{20}
\end{equation}
where
\begin{equation}
V(a)=-\frac{\kappa^2B_0^2a^2}{6\left(\beta B_0^2+a^4\right)}.
\label{21}
\end{equation}
Equation (20) is similar to an equation describing one-dimensional motion of a particle in the potential $V(a)$ with the energy $\epsilon$. The potential function (21) is negative and possesses a maximum at $a = a_c=\beta^{1/4}\sqrt{B_0}$. Let us consider the case of three-dimensional flat space, $\epsilon=0$. In this case, Eq. (20) can be integrated, and the result is
\begin{equation}
\sqrt{a^4+a_c^4}+a_c^2\ln\frac{a^2}{\sqrt{a^4+a_c^4}+a_c^2}
=\pm\left(\sqrt{\frac{2}{3\beta}}\kappa a_c^2t+C\right),
\label{22}
\end{equation}
where $C$ is a constant of integration. It is obvious to use the solution in Eq. (22) with the sign ($+$) in the right side of (22) when the scale factor increases in the time.
It should be noted that solution (22) for large $t$ gives, $a\propto \sqrt{t}$, the same behavior of the scale factor as for the radiation era in Maxwell's theory. In the early evolution for the small cosmic time, the dependence of the scale factor on the time becomes almost exponential as follows from Eq. (22) because nonlinear corrections to the Maxwell theory are essential.
The constant $C$ gives only the shift in the time and can be neglected.
The plot of the function $y=a/a_c$ vs $x=\sqrt{2}\kappa t/\sqrt{3\beta}$ for $C=0$ is given in Fig. 1.
\begin{figure}[h]
\includegraphics[height=4.0in,width=4.0in]{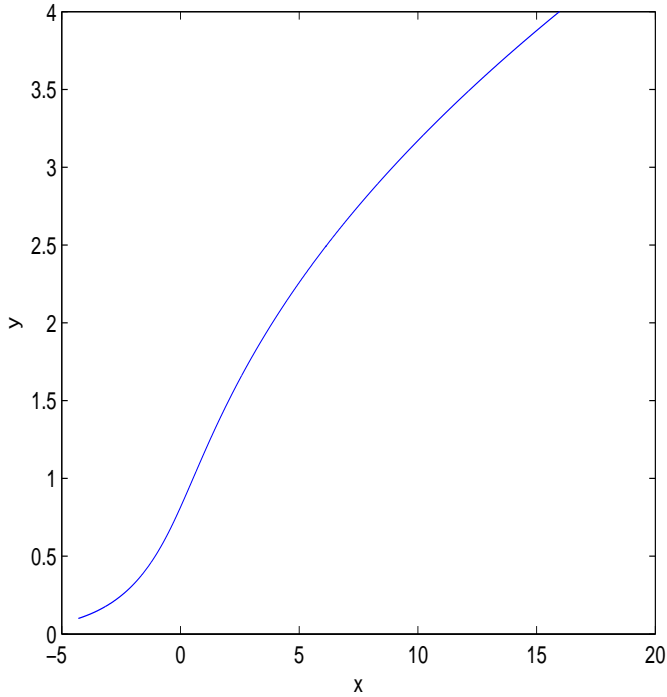}
\caption{\label{fig.1}The function  $y$  vs $x$.}
\end{figure}
At $t=0$ ($x=0$), we get from (22) [at $C=0$ and using $+$ in the right side of (22)] the equation for the radius of the universe $a_0$:
\begin{equation}
\sqrt{(a_0/a_c)^4+1}+\ln\frac{(a_0/a_c)^2}{\sqrt{(a_0/a_c)^4+1}+1}=0.
\label{23}
\end{equation}
The solution to the transcendental Eq. (23) is $a_0=0.814 a_c=0.814 \beta^{1/4}\sqrt{B_0}$. Thus, there is not singularity at the cosmic time $t=0$. As $a_0< a_c$, the universe undergos the acceleration expansion.
In accordance with the graph, the universe accelerates till the time when $y=1$ ($a=a_c$). At $y=1$, the acceleration stops, and the value $x\approx 0.533$ holds corresponding to the time $t\approx 0.65\sqrt{\beta}/\kappa$. After this time, the universe decelerates. Thus, in the toy model we suggest, one can explain inflation and the early time acceleration and avoid singularities at the beginning of the universe creation.

\subsection{Sound speed and causality}

One of the criteria for cosmological models to survive is that the speed of the sound is less that the local light speed, i.e. $c_s\leq 1$ \cite{Quiros}. Another criterion requires that the square sound speed is positive, $c^2_s> 0$. In this case there is no a classical instability. From Eqs. (4) and (5), we obtain the square sound speed (at $E=0$)
\begin{equation}
c^2_s=\frac{dp}{d\rho}=\frac{dp/d{\cal F}}{d\rho/d{\cal F}}=\frac{1-7\beta B^2}{3(\beta B^2+1)}.
\label{24}
\end{equation}
One can verify that the inequality $c_s\leq 1$ takes place ($\beta> 0$) for any value of the magnetic induction field $B$. Thus, the cosmological model under consideration does not admit superluminal fluctuations and meets the required bound $c_s\leq 1$. The requirement that the energy density perturbations do not uncontrollably grow leads to a classical stability, $c^2_s> 0$,
\begin{equation}
\frac{1-7\beta B_0^2/a^4}{3(\beta B_0^2/a^4+1)}> 0.
\label{25}
\end{equation}
According to Eq. (25), the scale factor has to obey the bound $a(t)>(7\beta)^{1/4}\sqrt{B_0}$ to possess a classical stability. At this stage, the universe undergoes the deceleration because the acceleration stops at $a_c=(\beta)^{1/4}\sqrt{B_0}$. As a result, we have a classical instability only during the period when $a(t)<(7\beta)^{1/4}\sqrt{B_0}$, i.e. during the inflation and a short time period of the universe deceleration. After, at $a(t)>(7\beta)^{1/4}\sqrt{B_0}$, the energy density perturbations do not grow.

\section{Conclusion}

A new model of nonlinear electromagnetic fields with a dimensional parameter $\beta$ coupled to GR, when the nonlinear electromagnetic field is the source of the gravitational field, has been considered. The scale invariance in this model is broken, and the trace of the energy-momentum tensor is nonzero. The nonlinear electromagnetic field possesses a finite value $E_{max}=1/\sqrt{\beta}$ at the origin of charged particles.
Some of the important problems in cosmology are the initial singularity and the accelerated expansion. We have considered  electromagnetic fields as a stochastic background with the nonzero value $<B^2>$.
It has been shown that the universe undergos inflation and the accelerated expansion. We show the absence of singularities at the beginning of the universe creation, in the energy density, in the curvature, in the Ricci tensor squared, and in the Kretschmann scalar. The magnetic universe is considered so that the magnetic field drives the universe to accelerate because we modified linear equations for the electromagnetic fields to be nonlinear equations. After some time, the universe decelerates, approaching the
Minkowski spacetime asymptotically. It was demonstrated that the speed of the sound is less than the local light speed; i.e. the causality takes place. A classical stability holds at the deceleration phase when $a(t)>(7\beta)^{1/4}\sqrt{B_0}$.
Homogeneous and isotropic cosmology with nonlinear electromagnetic radiation has been studied. Thus, in the cosmological model with NLE under consideration, the magnetic field allows the accelerated expansion of the universe. We presented here a model of NLE fields coupled to gravity which allows us to consider the universe evolution in accordance with the scenario of modern cosmology.

\vspace{5mm}
\textbf{Appendix}
\vspace{5mm}

Let us consider FRW spacetime with the flat spatial part with the line element
\begin{equation}
ds^2=-dt^2+a(t)\left(dx^2+dy^2+dz^2\right),
 \label{26}
\end{equation}
possessing the metric tensor elements
\begin{equation}
g_{00}=g^{00}=-1,~~g_{11}=g_{22}=g_{33}=a(t)^2,~~g^{11}=g^{22}=g^{33}=\frac{1}{a(t)^2},
 \label{27}
\end{equation}
where $a(t)$ is a scale factor. The standard calculations of nonzero components of the Christoffel symbols \cite{Landau}
\begin{equation}
\Gamma_{\mu\nu}^\alpha=\frac{1}{2}g^{\alpha\beta}\left(\frac{\partial g_{\beta \mu}}{\partial x^\nu}+
\frac{\partial g_{\beta \nu}}{\partial x^\mu}-\frac{\partial g_{\mu\nu}}{\partial x^\beta}\right)
\label{28}
\end{equation}
give ($\Gamma_{\mu\nu}^\alpha=\Gamma_{\nu\mu}^\alpha$)
\begin{equation}
\Gamma_{10}^1=\Gamma_{20}^2=\Gamma_{30}^3=\frac{\dot{a}}{a}=H,~~~~
\Gamma_{11}^0=\Gamma_{22}^0=\Gamma_{33}^0=\dot{a}a=a^2H,
\label{29}
\end{equation}
where $H=\dot{a}(t)/a(t)$ is the Hubble parameter.
The curvature tensor is given by \cite{Landau}
\begin{equation}
R_{~\nu\alpha\beta}^\mu=\frac{\partial \Gamma^\mu_{\nu\beta}}{\partial x^\alpha}-
\frac{\partial \Gamma^\mu_{\nu\alpha}}{\partial x^\beta}+\Gamma^\mu_{\sigma\alpha}\Gamma^\sigma_{\nu\beta}-
\Gamma^\mu_{\sigma\beta}\Gamma^\sigma_{\nu\alpha}
 \label{30}
\end{equation}
having the symmetry
\begin{equation}
R_{\mu\nu\alpha\beta}=-R_{\nu\mu\alpha\beta}=-R_{\mu\nu\beta\alpha},~~~~R_{\mu\nu\alpha\beta}=R_{\alpha\beta\mu\nu}.
 \label{31}
\end{equation}
From Eqs. (29) and (30), we find nonzero components of the curvature tensor
\[
R_{~00m}^i=-R_{~0m0}^i=\delta^i_m\frac{\ddot{a}}{a},~~R_{~k0m}^0=-R_{~km0}^0=\delta_{km}a\ddot{a},
\]
\begin{equation}
R_{klm}^i=\left(\delta^i_l\delta_{km}-\delta^i_m\delta_{kl}\right)\dot{a}^2.
\label{32}
\end{equation}
Then the Ricci tensor elements $R_{\nu\beta}= R_{~\nu\mu\beta}^\mu$ and a scalar curvature $R=R^\mu_{~\mu}$ read
\begin{equation}
R_{00}=-3\frac{\ddot{a}}{a},~~~~R_{ik}=\delta_{ik}\left(a\ddot{a}+2\dot{a}^2 \right),~~
R=6\left[\frac{\ddot{a}}{a}+\left(\frac{\dot{a}}{a}\right)^2\right]=\kappa^2(\rho-3p).
 \label{33}
\end{equation}
From Eq. (33) [see also Eqs. (9) and (19)], we obtain the squared of the Ricci tensor
\begin{equation}
R_{\mu\nu}R^{\mu\nu}=12\left[\left(\frac{\ddot{a}}{a}\right)^2+\frac{\ddot{a}}{a}\left(\frac{\dot{a}}{a}\right)^2 +\left(\frac{\dot{a}}{a}\right)^4\right]=\kappa^4\left(\rho^2+3p^2\right).
\label{34}
\end{equation}
One can calculate the Kretschmann scalar from Eqs. (32) [see also Eqs. (9) and(19)],
\begin{equation}
R_{\mu\nu\alpha\beta}R^{\mu\nu\alpha\beta}=12\left(\frac{\ddot{a}^2}{a^2}+\frac{\dot{a}^4}{a^4}\right)=
\kappa^4\left(\frac{5}{3}\rho^2+2\rho p+3p^2\right).
\label{35}
\end{equation}

\end{document}